\documentstyle[sprocl]{article}

\def\be{\begin{equation}}
\def\ee{\end{equation}}
\def\bea{\begin{eqnarray}}
\def\eea{\end{eqnarray}}

\begin{document}

\title{SOME EXACT SOLUTIONS OF THE DIRAC EQUATION}

\author{A.S. DE CASTRO}

\address{DFQ, UNESP, C.P. 205, 12500-000 Guaratinguet\'{a}
SP, Brasil\\E-mail: castro@feg.unesp.br}

\author{J. FRANKLIN}

\address{Department of Physics, Temple University, Philadelphia, PA
\\E-mail: v5030e@vm.temple.edu}

\maketitle

\abstracts{Exact analytic solutions are found to the Dirac equation for a
combination of Lorentz scalar and vector Coulombic potentials with
additional non-Coulombic parts. An appropriate linear combination of Lorentz
scalar and vector non-Coulombic potentials, with the scalar part dominating,
can be chosen to give exact analytic Dirac wave functions.} In a previous
letter,\cite{jf} simple exact solutions were found for the Dirac equation
for the combination of a Lorentz vector Coulomb potential with a linear
confining potential that was a particular combination of Lorentz scalar and
vector parts. In this work, we extend the method of Ref. 1 to the more
general case of an arbitrary combination of Lorentz scalar and vector
Coulombic potentials with a particular combination of Lorentz scalar and
vector non-Coulombic potentials. A more complete version of this work will
appear elsewhere.\cite{dcf}

The Dirac equation we solve is 
\begin{equation}
\left[{\bf \alpha\cdot p}+\beta m -\frac{\lambda+\beta\eta}{r} +V(r)+\beta
S(r)\right]\psi=E\psi,
\end{equation}
where {\boldmath$\alpha$\unboldmath} and $\beta$ are the usual Dirac
matrices. The four component wave function $\psi$ can be written in terms of
two component spinors $u$ (upper component) and $v$ (lower component)
satisfying the equations 
\begin{eqnarray}
({\bf \sigma\cdot p})v & = & \left[E-m+\frac{\lambda+\eta}{r}%
-V(r)-S(r)\right]u \\
({\bf \sigma\cdot p})u & = & \left[E+m+\frac{\lambda-\eta}{r}%
-V(r)+S(r)\right]v.
\end{eqnarray}
The key step in generating relatively simple exact solutions of the Dirac
equation is to choose a particularly simple form for the function $v(r)$.
For s-states,\footnote{States with orbital angular momentum are treated in
Ref.\ 2.} %
we choose $v(r)=i\gamma({\bf \sigma\cdot{\hat r}}%
)u(r)$, where $\gamma$ is a constant factor to be determined by the solution
to the Dirac equation. This is the form of $v(r)$ that was found in Ref.\ 1
for a Coulomb plus linear confining potential. This ansatz for $v(r)$ has
also been used as the basis for generating approximate saddle point
solutions for the Dirac\cite{sp1} and Breit\cite{sp2} equations. Using this
form of $v(r)$, equations (2) and (3) reduce to two first order ordinary
diffential equations for $u(r)$ 
\begin{eqnarray}
\frac{du}{dr} & = &\frac{1}{\gamma}\left[E-m-V(r)-S(r) +\frac{%
\lambda+\eta-2\gamma}{r} \right]u(r) \\
\frac{du}{dr} & = & -\gamma\left[E+m-V(r)+S(r)+\frac{\lambda-\eta}{r}%
\right]u(r).
\end{eqnarray}
Equations (4) and (5) are two independent equations for the same quantity,
so that each term in one equation can be equated with the corresponding term
in the other equation having the same radial dependence. This leads to 
\begin{equation}
\gamma^2=\frac{m-E}{m+E}=\frac{S(r)+V(r)}{S(r)-V(r)} =\frac{%
\lambda+\eta-2\gamma}{\eta-\lambda}.  \label{gs}
\end{equation}
The relations in Eq.\ (6) can be rearranged, after some algebra, to give 
\begin{equation}
\gamma=\frac{\lambda+\eta}{1+b},\;\;\; b=\pm\sqrt{1-\lambda^2+\eta^2}
\label{gb}
\end{equation}
The constant $b$ can have either sign. Although $b$ must be positive in the
pure Coulombic case, we will see that a negative $b$ is possible if the
Lorentz scalar potential $S(r)$ is more singular at the origin than $1/r$.
The bound state energy can be written as 
\begin{equation}
E=m\left(\frac{1-\gamma^2}{1+\gamma^2}\right) =m\left(\frac{b\lambda-\eta}{%
\lambda-b\eta}\right) =-m\frac{V(r)}{S(r)}.  \label{e}
\end{equation}
The wave function $u(r)$ can be found by solving Eq.\ (5) to give 
\begin{equation}
u(r)=r^{b-1}\exp\left[-a\left(r +\frac{1}{m}\int S(r)dr\right)\right],
\label{u}
\end{equation}
where the constant $a$ is given by $a=\gamma(m+E)=\pm\sqrt{m^2-E^2}$. The
constants $\gamma$ and $a$ can have either sign if $S(r)$ approaches a
constant or diverges as $r$ becomes infinite. The integral in Eq.\ (\ref{u})
can diverge for any finite $r$, as long as the product of $a$ times the
integral diverges in the positive sense. The integral can also diverge at
the origin or as $r\rightarrow\infty$ as long as the quantity in square
brackets in Eq.\ (\ref{u}) remains negative.

The last equality in Eq.\ (\ref{e}) shows that in order for this class of
exact solutions to apply, the Lorentz vector and scalar non-Coulombic
potentials must have the same radial dependence and opposite sign, with the
vector potential being smaller than the scalar potential. As long as this
constraint is satisfied, the results in Eqs.\ (\ref{gb})-(\ref{u}) represent
a complete exact solution for the ground state wave function and energy of
the Dirac Hamiltonian given in equation (1).

Because of the constraints imposed on the potentials by Eq.\ (\ref{e}), the
energy can be written purely in terms of one set of potentials or the other.
The ground state energy can be specified by the Coulombic constants $\lambda$
and $\eta$, with the other equality in Eq.\ (\ref{e}) serving as a
constraint on the non-Coulombic potentials $V(r)$ and $S(r)$. Or the ground
state energy could be specified by the ratio $V(r)/S(r)$ of the
non-Coulombic potentials with the other equality serving as a constraint on $%
\lambda$ and $\eta$. Although the possibility of this class of exact
solutions is limited by the constraints on the potentials, this still
permits a wide range of non-Coulombic potentials.

We now consider conditions imposed on the potentials and the wave function
parameters by the physical requirements that the potentials be real and the
wave function normalizable. We see from Eq.\ (\ref{gs}) that $\gamma$ must
be real, and then from Eq.\ (\ref{gb}) that $b$ must be real. This requires
the Coulombic potentials to satisfy the condition $1-\lambda^2+\eta^2 \ge 0$%
. The reality of $\gamma$ restricts possible bound state energies to the
range $-m < E < m$. Note that negative energies can occur, but $E+m$ cannot
be negative. Also, $E$ cannot equal $\pm m$, because this would make $%
\gamma=0$, leading to a constant, unormalizable wave function. This
condition on $E$, along with Eq.\ (\ref{u}), means that $V(r)$ must always
be less in magnitude than $S(r)$.

We discuss the remaining conditions on the parameters in terms of three
sub-classes of solution:

1. The ``normal" class of solutions has $b$, $\gamma $, and $a$ all
positive. In this case, we see from Eq.\ (\ref{gb}) that the Coulombic
potentials must satisfy the further condition $\lambda +\eta >0$.

2. This sub-case has $b$ negative, with $\gamma $ and $a$ still positive.
The constant $b$ can be negative if the product $aS(r)$ is positively
divergent at the origin faster than $1/r$. Then each of Eqs.\ (7)-(10) hold
just as for positive $b$, and the wave function is still normalizable. The
states with positive and negative $b$ are not two ground states of the same
Hamiltonian because the potentials cannot be the same for each state. That
is, either the Coulombic potentials or the non-Coulombic potentials must
change to be consistent with a negative $b$. Sub-case 1 with positive $b$
transforms smoothly into the pure Coulombic solution as the non-Coulombic
potential tends to zero everywhere. But this is not true for sub-case 2 with 
$b$ negative. This sub-case requires the non-Coulombic potential to be
dominant at the origin, and so has no corresponding pure Coulombic limit.

3. This sub-case has a negative $\gamma $ and a negative $a$, while $b$ can
have either sign, as discussed in sub-cases 1 and 2 above. A negative $a$ is
possible if the non-Coulombic potential diverges or approaches a constant as 
$r\rightarrow \infty $, so that the integral in Eq.\ (\ref{u}) diverges
faster than $r$ at large $r$. Since $a$ is negative, the potential $S(r)$
must be {\em negative} at large $r$. Then all of Eqs.\ (7)-(10) hold as for
positive $a$, and the wave function is still normalizable. This case is
highly unusual, because it allows the possibility of a potential that is
negative everywhere and diverges negatively at both the origin and infinite $%
r$. We know of no other example in quantum mechanics where such a negative
potential can lead to a normalizable ground state. The reason this is
possible here can be seen from Eqs.\ (4) and (5). There it is seen that $%
S(r) $ enters the differential equations for $u(r)$ only in the combinations 
$\gamma S(r)$ or $S(r)/\gamma $. Since these effective potentials are
positive, the resulting wave function is normalizable. As with sub-case 2,
the case with $\gamma $ and $a$ negative does not approach a pure Coulombic
case if the non-Coulombic potential tends to zero.

We now look at some special cases. If the non-Coulombic potentials are
absent, then the solutions are for a general linear combination of Lorentz
vector and Lorentz scalar Coulombic potentials. If either constant, $\lambda$
or $\eta$, is zero, we recover the usual solutions of the Dirac equation for
a pure scalar or vector Coulombic potential. The Coulombic potentials cannot
both be absent (while keeping a non-Coulombic part) because then $\gamma$
would be zero leading to a constant, unnormalizable wave function.

The method we have described does not work for radially excited states,
because the simple ansatz for $v(r)$ does not lead to consistent equations
for $du/dr$ in that case. The method does work for the lowest orbitally
excited state for which $l=j-\frac{1}{2}$. That case is discussed in Ref.\ 2.

\section*{Acknowledgments}

One of the authors (A.S.C.) would like thank FAPESP for financial support.

\end{document}